\documentclass[12pt,a4paper]{article}

\usepackage[utf8]{inputenc}
\usepackage[T1]{fontenc}
\usepackage{lmodern}
\usepackage[a4paper,margin=1in]{geometry}
\usepackage{amsmath,amssymb,mathtools}
\usepackage{graphicx}
\usepackage{booktabs}
\usepackage{siunitx}
\usepackage{microtype}
\usepackage{hyperref}
\usepackage[numbers,sort&compress]{natbib}

\hypersetup{
  colorlinks=true,
  linkcolor=black,
  citecolor=black,
  urlcolor=blue
}

\title{Urban Complexity through Vision Intelligence: Variance, Gradients, and Correlations across Six Italian Cities}

\author{
  Mirko Degli Esposti\textsuperscript{1,*} \and
  Armando Bazzani\textsuperscript{1} \and
  Chiara Dellacasa\textsuperscript{3} \and
  Matteo Falcioni\textsuperscript{1} \and
  Mario Massimo\textsuperscript{1} \and
  Martino Pietropoli\textsuperscript{2}\\[6pt]
  \textsuperscript{1}\,Department of Physics and Astronomy, University of Bologna, Italy\\
  \textsuperscript{2}\,Department of Architecture, University of Bologna, Italy\\
  \textsuperscript{3}\,CINECA, Bologna, Italy\\[4pt]
  \textsuperscript{*}\,Corresponding author: \href{mailto:mirko.degliesposti@unibo.it}{mirko.degliesposti@unibo.it}
}

\date{} 

\begin{document}
\maketitle

\begin{abstract}
This paper introduces a scalable methodology for the objective analysis of quality metrics across six major Italian metropolitan areas: Rome, Bologna, Florence, Milan, Naples, and Palermo. Leveraging georeferenced Street View imagery and an advanced Urban Vision Intelligence system, we systematically classify the visual environment, focusing on key metrics such as the Pavement Condition Index (PCI) and the Façade Degradation Score (FDS). The findings quantify \textit{Structural Heterogeneity (Spatial Variance)}, revealing significant quality dispersion (e.g., Milan $\sigma^2_{\mathrm{PCI}}=1.52$), and confirm that the classical \textit{Urban Gradient}---quality variation as a function of distance from the core---is consistently weak across all sampled cities ($R^2 < 0.03$), suggesting a complex, polycentric, and fragmented morphology. In addition, a \textit{Cross-Metric Correlation Analysis} highlights stable but modest interdependencies among visual dimensions, most notably a consistent positive association between façade quality and greenery ($\rho \approx 0.35$), demonstrating that structural and contextual urban qualities co-vary in weak yet interpretable ways. Together, these results underscore the diagnostic potential of Vision Intelligence for capturing the integrated spatial and morphological structure of Italian cities and motivate a large national-scale analysis.
\end{abstract}


% ...


\section{Introduction: Vision Intelligence and the Quantifiable City}

The visual dimension of cities has long been central to urban studies, from Lynch’s theory of \textit{imageability} to contemporary research on perception and aesthetics \cite{Zhang2024}. The convergence of large-scale geospatial imagery—particularly Google Street View (GSV)—and recent advances in Artificial Intelligence (AI) has enabled a new paradigm of \textit{Urban Visual Intelligence} (UVI) \cite{Fan2023, Zhang2024}. This framework allows the systematic observation and quantification of the built environment at the human scale, overcoming the limitations of traditional field surveys.

Recent research extends this paradigm through multimodal large language models (LLMs) and agentic reasoning for geospatial understanding. Models such as \textit{StreetViewLLM} \cite{Li2024}, \textit{StreetLens} \cite{Kim2025}, and \textit{UrbanSense} \cite{Yin2025} demonstrate how vision-language architectures can infer semantic and perceptual attributes directly from street-level imagery, while \textit{SAGAI} \cite{Perez2025} explores their generative potential to reconstruct and interpret urban scenes. In parallel, the new \textit{momepy.streetscape} module \cite{Araldi2025} provides a systematic framework for pedestrian-scale morphometrics. Collectively, these approaches signal the convergence of AI-based perception and quantitative morphology, forming the foundations of an emerging science of vision intelligence in the quantifiable city.

Beyond visual analysis, a complementary research trajectory envisions cities as ecosystems of interacting AI agents capable of reasoning, simulation, and goal-oriented behavior. Theoretical contributions such as \textit{Conceptualising the Emergence of Agentic Urban AI} \cite{Tiwari2025} and \textit{Towards Urban Planning AI Agent in the Age of Agentic AI} \cite{Fu2025} outline this shift from automation to agency, proposing frameworks in which autonomous agents participate in planning, governance, and adaptive decision-making. 
These perspectives extend the scope of urban AI from perception and representation to deliberation and action. 

Building on these advances, this paper introduces a high-resolution geospatial analysis aimed at quantifying infrastructure quality and spatial heterogeneity across six major Italian metropolitan areas—Rome, Bologna, Florence, Milan, Naples, and Palermo. The data were generated through \textit{UrbIA}, a multimodal agentic system that integrates language, vision, and spatial reasoning. Within this framework, 500 virtual agents per city—referred to as \textit{Humarels} (Human-scale Relational Agents)—simulate distributed visual observations of the urban environment. The complete architecture and operational deployment of the UrbIA system are detailed in a forthcoming paper \cite{UrbIA}.

Each Humarel samples multiple panoramic viewpoints through the GSV Static API, using controlled camera parameters to capture targeted urban metrics. For example, the \textit{Pavement Condition Index} (PCI) is obtained by setting the camera pitch between $\text{-35}^\circ$ and $\text{-45}^\circ$ to optimize the visibility of road surfaces.The imagery is then processed by a multimodal vision-language model performing automated visual assessment and scoring, producing georeferenced visual indicators such as surface quality, façade condition, and greenery presence.

The resulting dataset\footnote{Available upon request.} forms a consistent visual census for each city, from which we derive two primary indicators: the \textit{Spatial Variance}, capturing intra-urban heterogeneity, and the \textit{Urban Gradient}, describing the variation of visual quality with distance from the historical core. A further \textit{Cross-Metric Correlation Analysis} explores interdependencies among perceptual and infrastructural attributes, such as the relationship between pavement condition, greenery, and façade quality.

These layers of analysis—variance, gradient, and correlation—provide an integrated picture of urban quality, showing how structural and perceptual dimensions interact within complex, polycentric city structures. The results highlight the potential of vision-based intelligence systems for large-scale, comparative urban studies and support a broader research agenda toward explainable, AI-mediated urban analytics.

The remainder of the paper is organized as follows. Section~2 details the data acquisition and vision analysis pipeline. Section~3 presents comparative results across cities, focusing on spatial variance and urban gradients. Section~4 introduces the cross-metric correlation analysis, and Section~5 discusses city-specific patterns. Section~6 concludes with reflections on the implications of vision intelligence for urban analysis and planning.

\section{Materials and Methods}

Our approach integrates systematic data acquisition using the Google Maps Platform (GMP) APIs with a custom Vision Intelligence pipeline to produce a standardized, quantifiable representation of the urban environment.


The study encompasses six major Italian cities: \textit{Rome, Bologna, Florence, Milan, Naples, and Palermo}. These cities were selected to represent a diverse array of urban forms, population densities, and geographical contexts across the Italian peninsula.

In the initial development phase of the UrbIA  system, we explored two distinct sampling strategies for the deployment of the virtual agents (Humarels):
\begin{enumerate}
    \item \textit{Path-Based Sensing:} Utilizing the Directions API to define a continuous, high-density route (e.g., a critical urban corridor) and generating image captures every a 20 meters along the computed polyline. This method is effective for \textit{linear analysis} and capturing gradients along specific axes.
    \item \textit{Random Area Sampling (Adopted):} Selecting a fixed number of agents ($N=500$) based on random coordinates within the city's administrative area or a defined bounding box. This approach prioritizes  spatial coverage and statistical representativeness of the entire urban fabric, avoiding bias towards predefined routes.
\end{enumerate}
For the comparative analysis of the {\it Urban Gradient} and {\it Spatial Variance} presented in this paper, the {\it Random Area Sampling} method was utilized to ensure a statistically robust and unbiased assessment of the overall heterogeneity within each city's core region.

The area of study for each city was defined by its \textit{Administrative Bounding Box} (BB), typically encompassing the municipal or metropolitan boundary. Within this BB, we ensured the presence of available Street View imagery before selecting the $N=500$ random coordinates. The final sample size and the geographic coordinates of the study regions are provided in Table~\ref{tab:study_area}.

\begin{table}[htbp]
    \centering
    \caption{Study Areas Defined by Administrative Bounding Box}
    \label{tab:study_area}
    \begin{tabular}{l c c c}
        \toprule
        \textbf{City} & \textbf{BB Area ($\text{km}^2$)} & \textbf{Latitude Min/Max ($\circ$)} & \textbf{Longitude Min/Max ($\circ$)} \\
        \midrule
        Bologna & 140.73 & $44.40 / 44.57$ & $11.23 / 11.44$ \\
        Florence & 102.41 & $43.71 / 43.83$ & $11.12 / 11.31$ \\
        Milan & 181.76 & $45.40 / 45.54$ & $9.10 / 9.27$ \\
        Naples & 117.27 & $40.79 / 40.89$ & $14.18 / 14.34$ \\
        Palermo & 158.9 & $38.03 / 38.20$ & $13.24 / 13.43$ \\
        Rome & 1285.31 & $41.76 / 42.04$ & $12.33 / 12.68$ \\
        \bottomrule
    \end{tabular}
\end{table}


Each of the randomly selected Humarel locations serves as the point of origin for the image capture process. This step is critical as it transforms a single coordinate into a rich, multi-perspective visual dataset, maximizing the information gain from a static location.

Since the Humarel coordinates are randomly sampled and not path-dependent, the concept of a constant \texttt{travel\_heading} is irrelevant. Instead, we employ a {\it Rotational Sampling·} strategy to capture the full 360-degree environment visible from each point. For metrics requiring a full panoramic understanding (e.g., street width, context), multiple images are generated at fixed angular increments (e.g., $0^\circ, 90^\circ, 180^\circ, 270^\circ$ for a 4-view capture). This systematic rotation of the camera's {\it Heading} ensures that all adjacent facades, street corners, and contextual elements are documented.

The Street View Static API allows for precise control over the camera's orientation via the {\it Pitch} (vertical angle) and {\it Heading} (horizontal angle) parameters. This feature is exploited to perform {\it Targeted Vision Sensing}, isolating the specific urban feature required for the analysis. 

The final step in our methodology involves transforming the raw, targeted visual 
data generated by the Humarels into quantifiable urban metrics. For the analysis 
presented in this paper, we adopt a direct prompt-based scoring approach using 
the GPT-4 Vision multimodal language model. 

For each metric—Pavement Condition Index (PCI), Façade Degradation Score (FDS), 
Green Presence, Graffiti Index, and Urban Canyon Index—we designed specific 
text prompts that instruct the model to evaluate the street-level image and 
return a numerical score within a predefined range (e.g., 1–5 for PCI and FDS). 
This prompt engineering strategy leverages the zero-shot and few-shot reasoning 
capabilities of vision-language models to produce direct quality assessments 
without requiring explicit pixel-level segmentation or custom-trained classifiers.

We acknowledge that this approach, while straightforward and scalable, represents 
a preliminary implementation of vision-based urban sensing. The reliance on 
holistic image-level scoring rather than fine-grained spatial analysis introduces 
limitations in interpretability and geometric precision. However, the consistency 
of results across cities and the statistical robustness of derived indicators 
suggest that prompt-based scoring captures meaningful urban quality patterns at 
scale.\footnote{Our broader research program (UrbIA \cite{UrbIA}) integrates 
dedicated geometric segmentation models (e.g., SA2VA \cite{Yuan2025}) for 
precise pixel-level analysis, quantifying the percentage of road surface, greenery, 
sky, graffiti, and defects. These more sophisticated vision architectures will be 
deployed in future large-scale studies. The results presented here serve to 
establish baseline metrics and validate the feasibility of vision intelligence for 
comparative urban analysis.}

The output of this comprehensive pipeline is a georeferenced database containing the objective visual score for each Humarel location across the six cities, which forms the empirical basis for the gradient and variance analysis presented in the following sections.

\subsection{Illustrative Humarel Observations}

To provide a qualitative overview of the visual content processed by the Humarel agents, this subsection presents a selection of representative Street View frames acquired during the data collection phase. Each Humarel captures multiple perspectives of its surroundings by varying the camera's heading and pitch, as described in Section~2. The images shown here were intentionally selected to illustrate clear examples of the visual features used to compute the metrics introduced in this paper.

\begin{figure}[htbp]
    \centering
    \includegraphics[width=\textwidth]{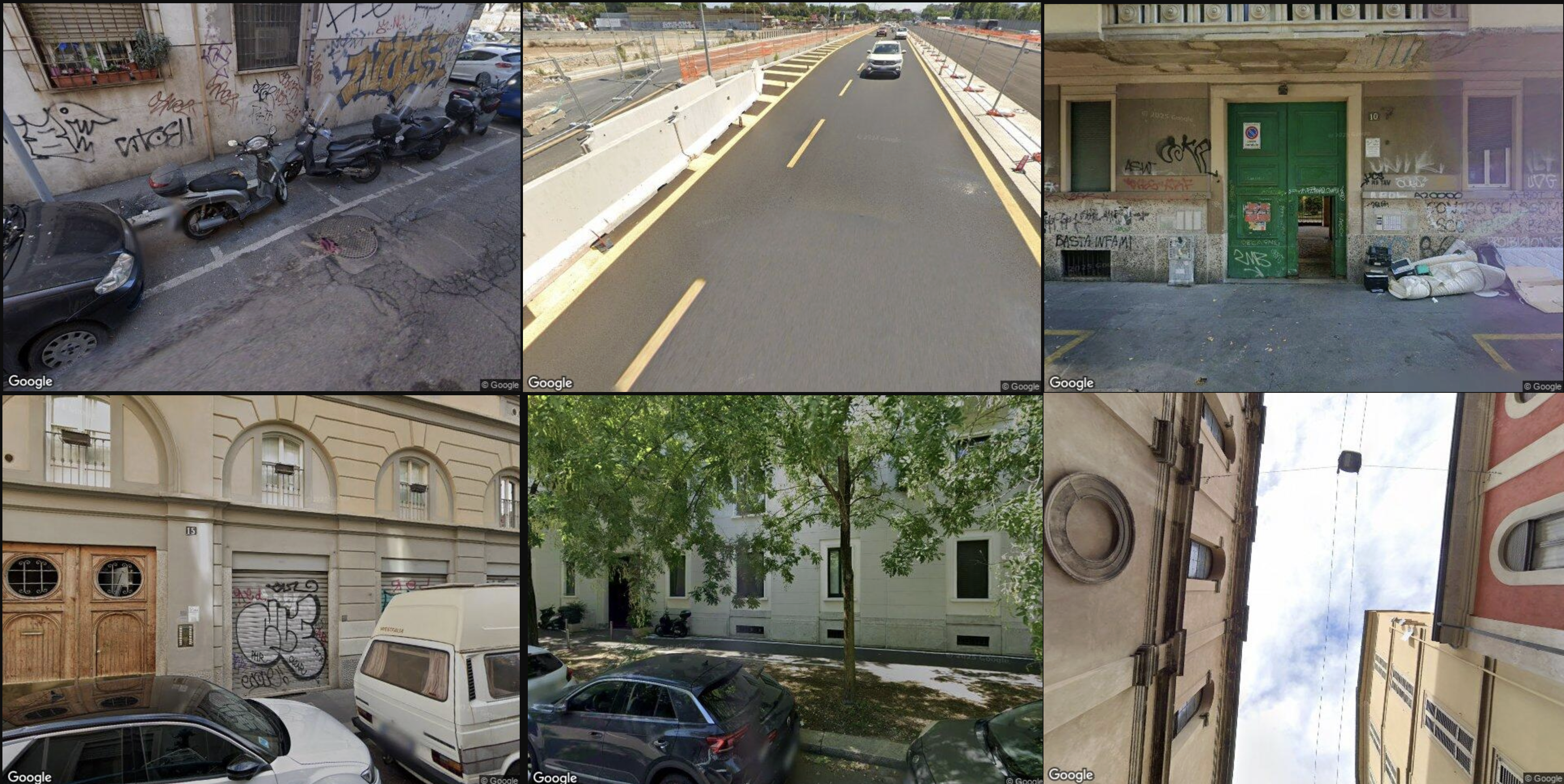}
    \caption{Representative examples of the visual content captured by Humarel agents across the six cities. 
    The top row illustrates structural conditions, including pavement and façade degradation, while the bottom row shows contextual and morphological features such as graffiti, greenery, and the urban canyon effect. 
    Images are included solely for qualitative illustration of the UrbIA vision sensing process.}
    \label{fig:humarel_examples}
\end{figure}

Figure~\ref{fig:humarel_examples} shows several representative examples of the targeted vision sensing used to derive both structural and contextual indicators. The top row includes cases of \textit{pavement degradation} and \textit{façade deterioration}, corresponding respectively to low Pavement Condition Index (PCI) and low Façade Degradation Score (FDS) values. The bottom row presents additional examples illustrating \textit{graffiti presence}, \textit{urban greenery}, and the \textit{urban canyon effect}, where tall façades and narrow street sections define highly enclosed spatial configurations.

While the examples in Figure~\ref{fig:humarel_examples} were manually chosen for clarity, in practice the dataset also contains a fraction of frames that are visually ambiguous or partially unusable due to occlusions, strong shadows, or camera artefacts. A preliminary heuristic inspection suggests that such anomalous or low-quality images account for approximately 15\% of the total sample. This issue is intrinsic to large-scale image-based urban sensing and will require further refinement of our filtering and quality-control procedures. Nevertheless, we emphasize that the statistical indicators introduced in this study---including the variance, gradient, and correlation measures---are designed to be robust to this level of noise, as confirmed by their stability and consistency across all six metropolitan areas.

\begin{table}[htbp]
    \footnotesize 
    \centering
    \caption{Street View Camera Parameters}
    \label{tab:pitch_heading}
    \begin{tabular}{p{3.0cm} cc p{3.5cm}}
        \toprule
        \textbf{Metric Target} & \textbf{Pitch (V.)} & \textbf{Heading (H.)} & \textbf{Analytical Focus} \\
        \midrule
        \textbf{Pavement Condition (PCI)} & $-35^{\circ}$ or $-45^{\circ}$ & Aligned to street segment & Road surface wear, material defects (\textit{Manto}). \\
        \textbf{Façade Degradation (FDS)} & $0^{\circ}$ (Horizon) & Rotational Sampling & Building quality and deterioration (\textit{Facade}). \\
        \textbf{Green Presence Index} & $0^{\circ}$ (Horizon) & Rotational Sampling & Visible vegetation density and canopy cover (\textit{Verde}). \\
        \textbf{Urban Canyon Index} & $0^{\circ}$ (Horizon) & Rotational Sampling & Street enclosure, SVF surrogate (\textit{Canyon}). \\
        \textbf{Graffiti Index} & $0^{\circ}$ (Horizon) & Rotational Sampling & Presence and extent of visible graffiti (\textit{Graffiti}). \\
        \textbf{Sky View Factor (SVF)} & $+90^{\circ}$ (Zenith) & Irrelevant & Microclimatic analysis, visible sky measurement. \\
        \bottomrule
    \end{tabular}
        \label{tab:targeted_sensing_updated}
\end{table}

\section{Results: Quantifying Urban Disparities}


The analysis of the Humarel data across the six metropolitan areas yielded a rich dataset of street-level metrics. We emphasize that the results presented in this section are preliminary findings derived from the initial $N=500$ random samples per city. These results serve primarily to demonstrate the predictive power and scalability of the UrbIA Vision Intelligence approach. While the trends observed are statistically significant within the sampled population, future work will involve refining the Vision AI models and expanding the sample size to validate these findings against established socioeconomic and municipal infrastructure data.


The urban quality metrics extracted by the \textit{UrbIA Multimodal Vision Agents} were analyzed by grouping them into two categories: \textit{Structural Quality Metrics} (Pavement Condition Index — PCI, and Façade Degradation), for which both the mean value and the Spatial Variance ($\sigma$, $\sigma^2$) are key indicators, and \textit{Contextual Metrics} (Aesthetic and Morphological). Table~\ref{tab:structural_quality_stats} reports the updated statistics computed from the full Humarel dataset for the six metropolitan areas.

\begin{table}[htbp]
    \centering
    \small
    \caption{Structural Quality Metrics: Mean Scores, Standard Deviation ($\sigma$), and Variance ($\sigma^2$) computed on 500 Humarel points per city.}
    \label{tab:structural_quality_stats}
    \begin{tabular}{l c c c c c c}
        \toprule
        \textbf{City} & \textbf{Mean PCI} & \textbf{PCI $\sigma$} & \textbf{PCI $\sigma^2$} & \textbf{Mean Façade} & \textbf{Façade $\sigma$} & \textbf{Façade $\sigma^2$} \\
        & (1–5) & & & (1–5) & & \\
        \midrule
        Bologna & 3.37 & 0.99 & 0.99 & 3.55 & 1.03 & 1.07 \\
        Florence & 3.21 & 0.97 & 0.94 & \textbf{4.26} & 1.00 & 1.00 \\
        Milan & 3.22 & \textbf{1.23} & \textbf{1.52} & 3.63 & \textbf{1.09} & \textbf{1.19} \\
        Naples & 3.18 & 0.82 & 0.68 & 2.92 & 0.94 & 0.89 \\
        Palermo & \textbf{2.93} & \textbf{0.69} & \textbf{0.48} & 3.07 & 0.77 & 0.60 \\
        Rome & 3.08 & 1.00 & 1.01 & 3.44 & 1.01 & 1.03 \\
        \bottomrule
    \end{tabular}
    \vspace{0.5ex}
    \begin{flushleft}
    \footnotesize \textit{Note:} Bold values indicate extreme values (highest or lowest) within each column.
    \end{flushleft}
\end{table}

\begin{table}[htbp]
    \centering
    \small 
    \caption{Contextual Metrics: Average Scores (Aesthetic and Morphological)}
    \label{tab:contextual_stats}
    \begin{tabular}{l c c c}
        \toprule
        \textbf{City} & \textbf{Graffiti Index} & \textbf{Green Presence} & \textbf{Urban Canyon} \\
        & (0-2) & (0-5) & (0-2) \\
        \midrule
        Bologna & 0.48 & 1.99 & 0.84 \\
        Florence & \textbf{0.13} & \textbf{2.75} & 1.30 \\
        Milan & 0.36 & 2.30 & 0.90 \\
        Naples & 0.45 & \textbf{1.72} & 1.08 \\
        Palermo & 0.30 & 2.40 & \textbf{1.43} \\
        Rome & \textbf{0.48} & 2.73 & 1.20 \\
        \bottomrule
    \end{tabular}
    \vspace{1ex} 
   
\end{table}


The comparison of average scores reveals distinct urban profiles across the six cities:

\begin{itemize}
\item \textit{Pavement Condition (PCI):} \textbf{Bologna} and \textbf{Milan} exhibit the highest average pavement scores ($\mathbf{3.37}$ and $\mathbf{3.22}$, respectively), indicating comparatively better road surface conditions. \textbf{Palermo}, with the lowest mean PCI ($\mathbf{2.93}$), confirms the greatest overall need for maintenance intervention. In terms of spatial variance, \textbf{Milan} shows the highest dispersion ($\sigma^2 = 1.52$), pointing to pronounced heterogeneity between well-maintained and deteriorated areas, while \textbf{Palermo} displays the lowest variance ($\sigma^2 = 0.48$), suggesting more uniformly modest conditions across the urban fabric.

\item \textit{Façade Condition (FDS):} \textbf{Florence} stands out with the highest average façade score ($\mathbf{4.26}$), reflecting stronger preservation practices and visual consistency of its built environment. Conversely, \textbf{Naples} records the lowest mean façade score ($\mathbf{2.92}$), indicating widespread degradation. Regarding variance, \textbf{Milan} again exhibits the highest dispersion ($\sigma^2 = 1.19$), consistent with a visually diverse building stock, whereas \textbf{Palermo} shows the lowest façade variance ($\sigma^2 = 0.60$), highlighting more homogeneous, though moderately degraded, façades.

    \item \textit{Green Presence:} \textbf{Florence} ($\mathbf{2.75}$) and \textbf{Rome} ($\text{2.73}$) exhibit the highest average scores for visible green presence. \textbf{Naples} recorded the lowest score ($\mathbf{1.72}$), indicating a low density of visible vegetation in the Humarels' viewpoints.
    
    \item \textit{Graffiti Index:} The average presence of visible graffiti is highest in \textbf{Bologna} and \textbf{Rome} ($\mathbf{0.48}$). Notably, \textbf{Florence} recorded the lowest average index ($\mathbf{0.13}$), suggesting effective cleaning or a lower incidence within its core study area.
    
    \item \textit{Urban Canyon Index:} This index, measuring the perceived enclosure of the street space, is highest in \textbf{Palermo} ($\mathbf{1.43}$), indicating a prevalence of extremely narrow streets and dense building morphology. \textbf{Bologna} shows the lowest index ($\mathbf{0.84}$), reflecting a sampling dominated by more open avenues or boulevards.
\end{itemize}


    



The analysis of simple average scores, while useful for macro-level comparison (Section 3.2), fails to capture the defining characteristic of the Italian urban experience: {\it extreme spatial heterogeneity}. Unlike many modern cities where quality often follows predictable sectorial or radial patterns, historic Italian cities exhibit high contrast and rapid transitions in urban quality. It is common to find beautifully preserved areas—high-scoring tourist centers—immediately adjacent to segments displaying significant degradation in terms of pavement condition or façade maintenance, often mere meters away (the "degrado dietro l'angolo" effect). This phenomenon of {juxtaposition of excellence and deficit} is a critical dimension of urban resilience and planning efficiency.


\subsection{Spatial Variance (Heterogeneity)}
Spatial Variance quantifies the degree of disparity or non-uniformity of a given metric within the boundaries of a single city. A high variance in a score (e.g., Pavement Condition Index) indicates that the city is deeply heterogeneous, suggesting a significant difference in infrastructure quality between the best-maintained and the worst-maintained areas. This can be calculated using the standard deviation ($\sigma$) or the coefficient of variation (CV) of the 500 Humarels' scores for each metric.

\begin{figure}[htbp]
    \centering
    \begin{minipage}[t]{0.48\textwidth}
        \centering
        \includegraphics[width=\textwidth]{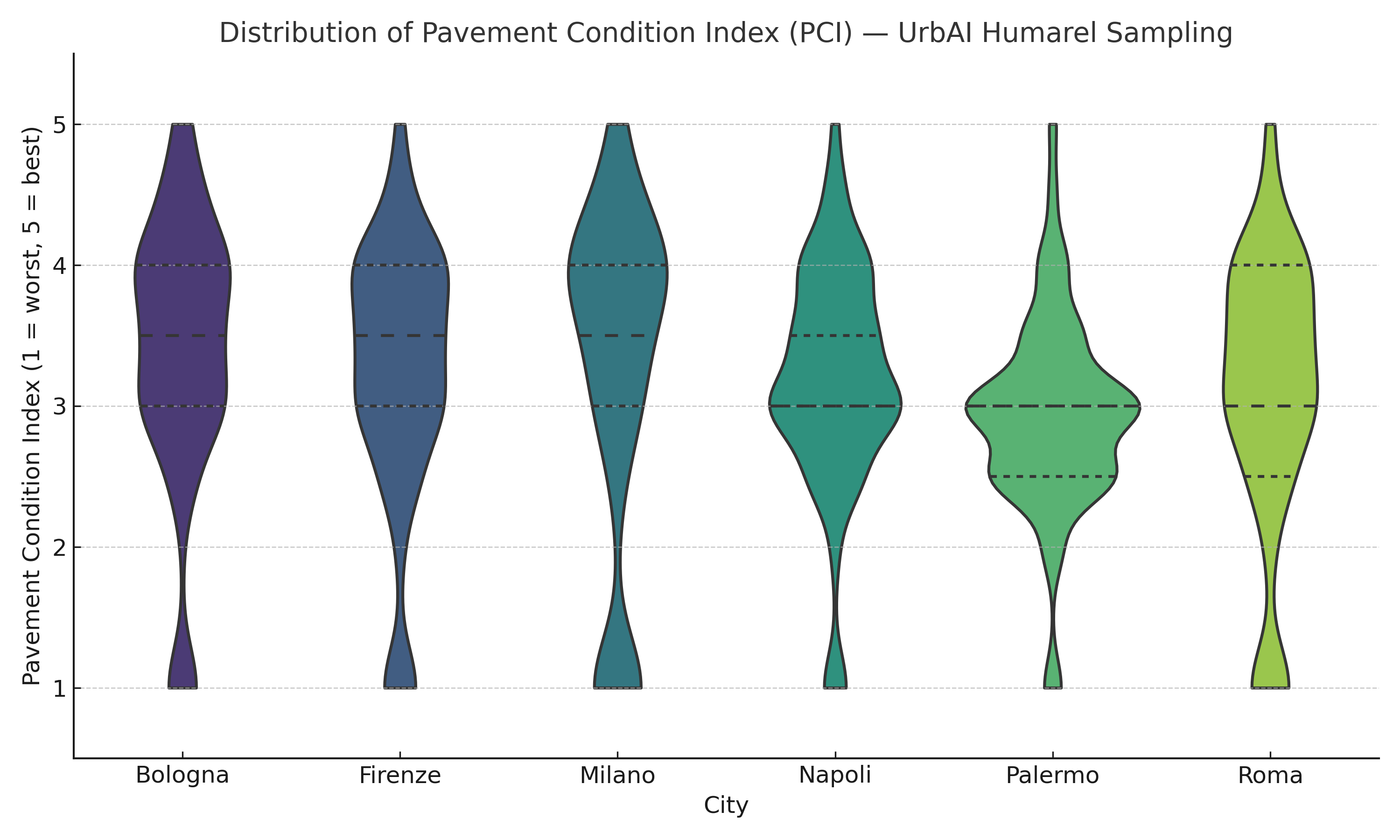}
        \smallskip
        \small{(\textbf{a}) Pavement Condition Index (PCI)}
        \label{fig:violin_pci}
    \end{minipage}
    \hfill
    \begin{minipage}[t]{0.48\textwidth}
        \centering
        \includegraphics[width=\textwidth]{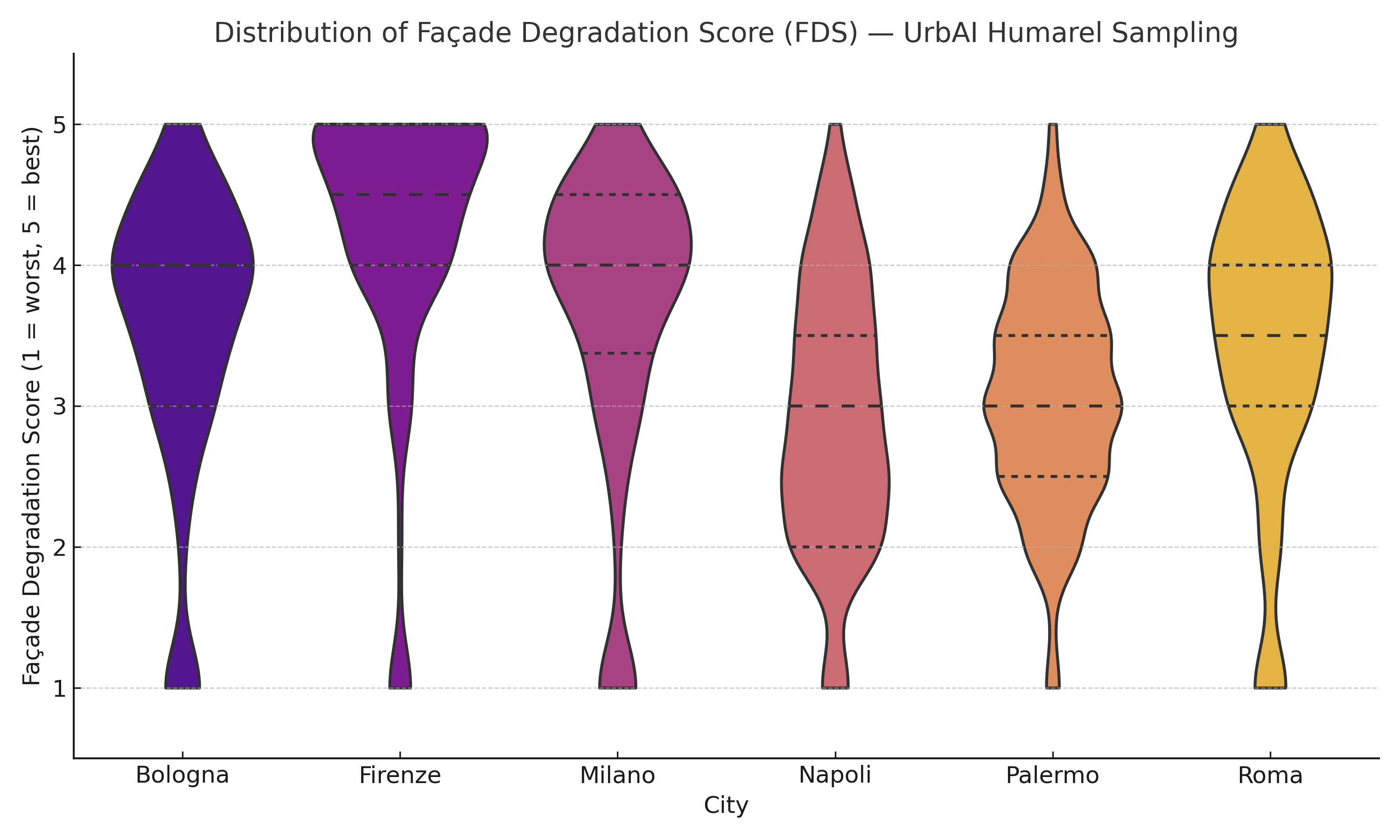}
        \smallskip
        \small{(\textbf{b}) Façade Degradation Score (FDS)}
        \label{fig:violin_fds}
    \end{minipage}
    \caption{
        Empirical distributions of structural visual quality scores across six Italian metropolitan areas (Humarel sampling, $N{=}500$ per city). 
        Violin plots display the full density of scores (1--5), with medians and extrema. 
        PCI and FDS share the same y-axis range for comparability. 
        The plots complement Table~\ref{tab:structural_quality_stats} by revealing the internal shape of spatial heterogeneity 
        (e.g., skewness, multimodality), thereby motivating the Urban Gradient analysis introduced in the next section.
    }
    \label{fig:violin_structural}
\end{figure}

Figure~\ref{fig:violin_structural} visualizes the empirical distributions of the Pavement Condition Index (PCI) and the Façade Degradation Score (FDS) for the six metropolitan areas. Each violin shows the density of Humarel scores, with median and interquartile markers. The figure complements Table~\ref{tab:structural_quality_stats} by illustrating the internal shape of spatial heterogeneity: Milan and Bologna display broad PCI distributions with visible bimodality, while Palermo and Naples show narrower, lower-centered profiles consistent with their lower means. In façade quality, Florence presents a compact, upper-shifted distribution (homogeneously high-quality façades), whereas Naples exhibits a long lower tail, indicating spatially clustered degradation. These distributions confirm that visual quality is not evenly distributed within cities, providing the empirical rationale for the Urban Gradient analysis discussed in the following section.


\subsection{ the Urban Gradient: Center Selection}

The {\it Urban Gradient} quantifies how visual quality varies as a function of distance from the historical core of the city. For each Humarel observation $i$, the gradient is modeled as the relationship between the measured visual score (e.g., the Pavement Condition Index, PCI, or the Façade Degradation Score, FDS) and the {\it Distance from the Historical Center} (DHC), expressed in kilometers.

A critical methodological step is the definition of the anchor point for DHC. Italian cities are often polycentric or possess layered historical centers that complicate the notion of a single ``city center.'' To ensure comparability and reproducibility, we selected for each metropolitan area a landmark that is both historically and functionally central—typically the main cathedral square or administrative nucleus.

Table~\ref{tab:city_centers} lists the selected anchor points used for calculating DHC for all 500 Humarel observation points per city. Geographic distances were computed using the Haversine formula on latitude–longitude coordinates, providing great-circle distances in kilometers.

\begin{table}[htbp]
    \centering
    \caption{Anchor Points for Distance-from-Historical-Center (DHC) Calculation. Coordinates are expressed in decimal degrees.}
    \label{tab:city_centers}
    \begin{tabular}{llcc}
        \toprule
        \textbf{City} & \textbf{Center Location Used} & \textbf{Latitude} & \textbf{Longitude} \\
        \midrule
        Bologna & Piazza Maggiore & $44.4938^\circ$ N & $11.3426^\circ$ E \\
        Florence & Piazza del Duomo & $43.7730^\circ$ N & $11.2561^\circ$ E \\
        Milan & Piazza del Duomo & $45.4642^\circ$ N & $9.1900^\circ$ E \\
        Naples & Piazza del Plebiscito & $40.8384^\circ$ N & $14.2494^\circ$ E \\
        Palermo & Quattro Canti & $38.1147^\circ$ N & $13.3619^\circ$ E \\
        Rome & Piazza del Campidoglio & $41.8933^\circ$ N & $12.4828^\circ$ E \\
        \bottomrule
    \end{tabular}
\end{table}

The Urban Gradient is operationally defined as the slope $\beta_1$ in a linear model of the form:
\[
S_i = \beta_0 + \beta_1 \, DHC_i + \varepsilon_i,
\]
where $S_i$ represents either PCI or FDS for observation $i$, and $DHC_i$ is the corresponding distance from the city’s historical center. Negative values of $\beta_1$ indicate a decline in visual quality with increasing distance from the historical core.

By fitting regression models across the six cities, the gradient analysis enables us to:
\begin{enumerate}
    \item Evaluate whether the classical concentric urban model holds—i.e., if visual quality decays radially with distance.
    \item Identify non-linear or segmented trends in the quality–distance relationship, characteristic of polycentric, industrial, or postmodern urban morphologies \cite{Zhang2024}.
\end{enumerate}

\begin{figure}[htbp]
    \centering
    \includegraphics[width=\textwidth]{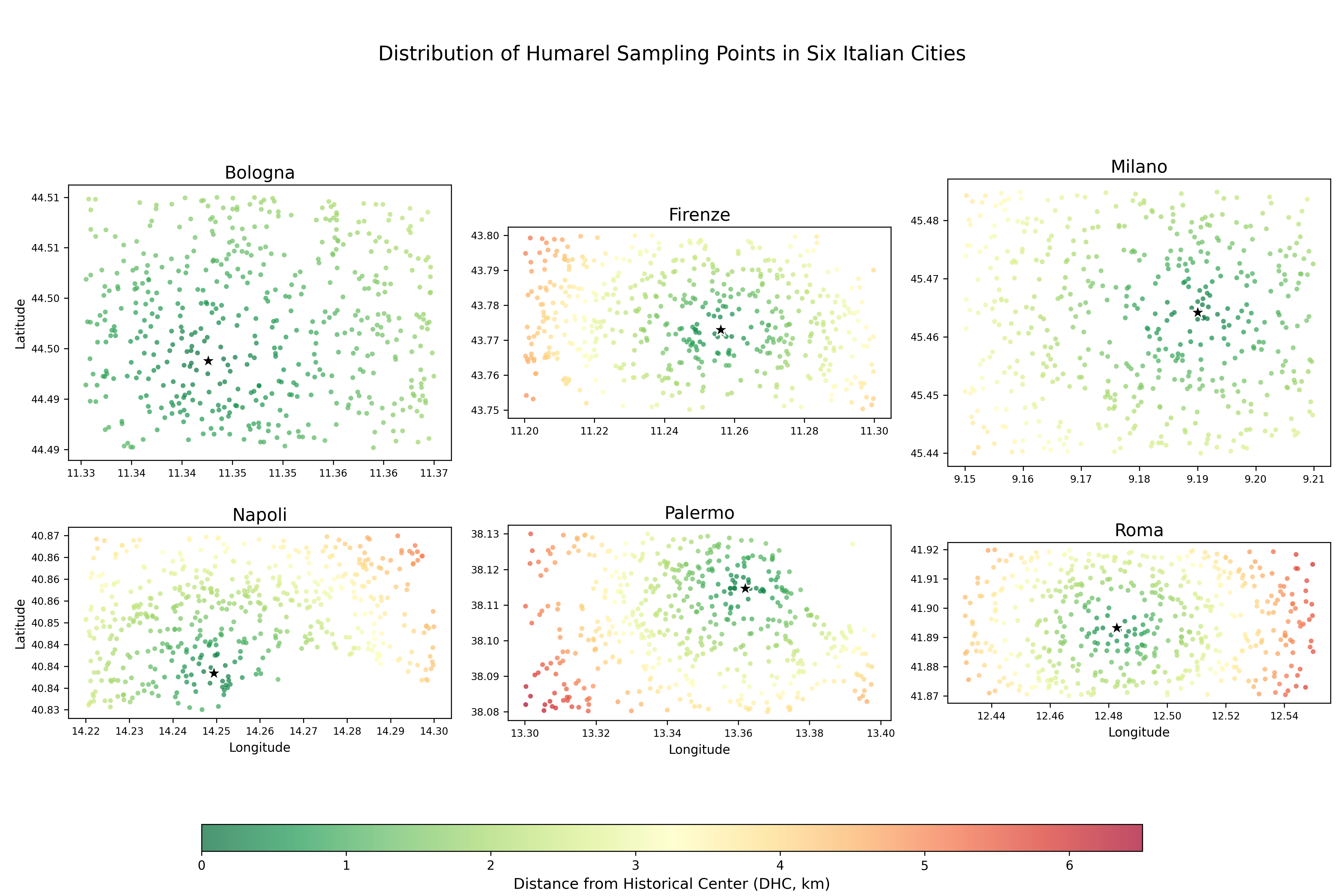} 
    \caption{Distribution of the 500 Humarel Sampling Points in the six study cities. The location of the Historical Center (anchor point for DHC) is marked with a distinctive symbol.}
    \label{fig:city_maps}
\end{figure}

To visualize the spatial coverage and distribution of the 500 Humarel sampling points across the study areas, Figure~\ref{fig:city_maps} presents a static map for each city, with the points plotted relative to the Historical Center (Hollow Circle Marker), reflecting the shape of the city.

\subsection{Urban Gradient Results and Interpretation}

Table~\ref{tab:urban_gradient_results} reports the linear regression coefficients of the Urban Gradient model for both the Pavement Condition Index (PCI) and the Façade Degradation Score (FDS). Across the six Italian metropolitan areas, the slopes ($\beta_1$) are generally small, with $R^2 < 0.03$, indicating that radial distance from the historical center explains only a limited portion of the spatial variance. \textbf{Bologna}, \textbf{Milan}, and \textbf{Palermo} exhibit mildly negative PCI gradients, suggesting a gradual decline in pavement quality outward from the core, while \textbf{Florence}, \textbf{Bologna}, and \textbf{Palermo} show positive FDS gradients, reflecting newer and better-maintained façades in suburban districts. \textbf{Naples} and \textbf{Rome} display weak or contrasting patterns, consistent with their polycentric and historically layered structure.

\begin{table}[htbp]
    \centering
    \footnotesize
    \caption{Summary of Urban Gradient regressions: slope ($\beta_1$), determination coefficient ($R^2$), and statistical significance ($p$). Interpretations highlight the direction and strength of each metric's relationship with distance from the historical center (DHC).}
    \label{tab:urban_gradient_summary}
    \begin{tabular}{l l r r r p{7cm}}
        \toprule
        \textbf{City} & \textbf{Metric} & \textbf{$\beta_1$} & \textbf{$R^2$} & \textbf{$p$} & \textbf{Interpretation} \\
        \midrule
        Bologna & PCI & $-0.10$ & 0.003 & 0.23 & Very weak negative slope; not significant. \\
        & FDS & $+0.19$ & 0.009 & \textbf{0.03} & Significant positive trend; façades improve outward. \\
        \addlinespace
        Florence & PCI & $-0.02$ & 0.001 & 0.56 & Flat; no measurable gradient. \\
        & FDS & $+0.14$ & \textbf{0.025} & \textbf{< 0.001} & Strong positive; façade quality increases outward. \\
        \addlinespace
        Milan & PCI & $-0.09$ & 0.004 & 0.17 & Weak decline; not significant. \\
        & FDS & $+0.02$ & < 0.001 & 0.79 & Flat; no correlation. \\
        \addlinespace
        Naples & PCI & $0.00$ & < 0.001 & 0.99 & No gradient; uniform quality. \\
        & FDS & $-0.11$ & 0.016 & \textbf{0.004} & Negative slope; façade degradation increases outward. \\
        \addlinespace
        Palermo & PCI & $-0.07$ & \textbf{0.022} & \textbf{< 0.001} & Significant negative; quality declines outward. \\
        & FDS & $+0.03$ & 0.003 & 0.22 & Weakly positive; not significant. \\
        \addlinespace
        Rome & PCI & $+0.02$ & 0.001 & 0.53 & No meaningful gradient. \\
        & FDS & $-0.06$ & 0.006 & 0.08 & Slight decline; marginally significant. \\
        \bottomrule
    \end{tabular}
\end{table}

\noindent
These results confirm that radial distance alone explains only a minor share of the spatial variance in visual quality ($R^2 < 0.03$). The Urban Gradient analysis reveals weak or city-specific trends—negative PCI gradients in \textbf{Bologna}, \textbf{Milan}, and \textbf{Palermo}, and positive FDS gradients in \textbf{Florence}, \textbf{Bologna}, and \textbf{Palermo}—highlighting the complexity and polycentric nature of Italian urban morphologies. Such differentiated and metric-dependent patterns demonstrate the diagnostic value of the Urban Gradient framework and motivate its extension to a large-scale, data-intensive analysis.


\begin{table}[htbp]
    \centering
    \small
    \caption{Linear regression coefficients of the Urban Gradient model: $S_i = \beta_0 + \beta_1 DHC_i + \varepsilon_i$, for Pavement Condition Index (PCI) and Façade Degradation Score (FDS).}
    \label{tab:urban_gradient_results}
    \begin{tabular}{l l r r r r r}
        \toprule
        \textbf{City} & \textbf{Metric} & \textbf{$\beta_0$} & \textbf{$\beta_1$} & \textbf{$R^2$} & \textbf{$p$-value} & \textbf{$N$} \\
        \midrule
        Bologna & PCI & 3.49 & $-0.10$ & 0.003 & 0.23 & 500 \\
        Bologna & FDS & 3.33 & $+0.19$ & 0.009 & 0.031 & 500 \\
        Florence & PCI & 3.27 & $-0.02$ & 0.001 & 0.56 & 500 \\
        Florence & FDS & 3.90 & $+0.14$ & \textbf{0.025} & \textbf{< 0.001} & 500 \\
        Milan & PCI & 3.39 & $-0.09$ & 0.004 & 0.17 & 500 \\
        Milan & FDS & 3.60 & $+0.02$ & < 0.001 & 0.79 & 500 \\
        Naples & PCI & 3.17 & $0.00$ & < 0.001 & 0.99 & 500 \\
        Naples & FDS & 3.18 & $-0.11$ & 0.016 & 0.004 & 500 \\
        Palermo & PCI & 3.13 & $-0.07$ & \textbf{0.022} & \textbf{< 0.001} & 500 \\
        Palermo & FDS & 2.99 & $+0.03$ & 0.003 & 0.22 & 500 \\
        Rome & PCI & 3.01 & $+0.02$ & 0.001 & 0.53 & 500 \\
        Rome & FDS & 3.62 & $-0.06$ & 0.006 & 0.08 & 500 \\
        \bottomrule
    \end{tabular}
    \vspace{1ex}
\end{table}

\begin{figure*}[htbp]
    \centering
    \includegraphics[width=0.95\textwidth]{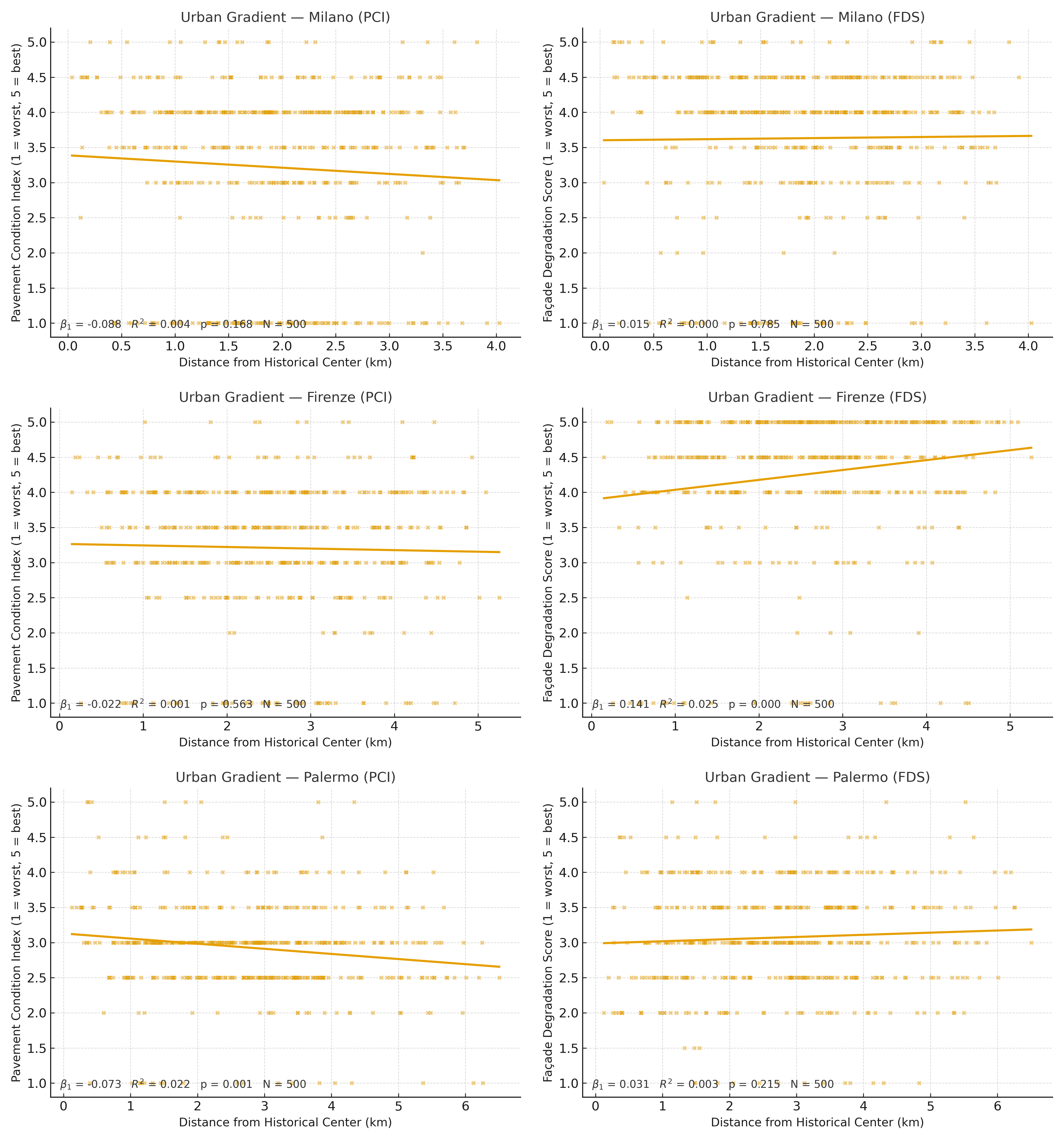}
    \caption{Representative Urban Gradient plots for three Italian metropolitan areas (Milan, Florence, and Palermo). Each scatter plot shows individual Humarel observations (n=500 per city) with fitted regression lines for both Pavement Condition Index (PCI, left column) and Façade Degradation Score (FDS, right column). Despite mild positive or negative slopes, the overall dispersion remains high, confirming that distance from the historical core explains only a small share of the observed spatial variance.}
    \label{fig:urban_gradient_examples}
\end{figure*}

Although the regression coefficients in Table~\ref{tab:urban_gradient_results} quantify the general direction and strength of the Urban Gradient, their small magnitudes do not fully convey the spatial structure of the underlying data. Figure~\ref{fig:urban_gradient_examples} visualizes representative scatter plots of Pavement Condition Index (PCI) and Façade Degradation Score (FDS) against the Distance from Historical Center (DHC). The plots confirm that while weak negative or positive trends exist, the dispersion of points is high and city-specific patterns diverge markedly. These visualizations illustrate the limited explanatory power of purely radial models and motivate the extension of this framework to non-linear and multi-center gradient analysis using large-scale computational modeling.

\subsection{Cross-Metric Correlation Analysis}

To complement the spatial and radial analyses presented in the previous sections, we examined the internal relationships among the structural and contextual visual metrics extracted by the UrbIA Multimodal Vision Agents. The goal of this correlation analysis is to determine whether specific aspects of the urban scene—such as pavement quality, façade condition, greenery, or graffiti presence—tend to co-vary across the 3,000 Humarel observation points collected from the six metropolitan areas.

Table~\ref{tab:correlation_matrix} reports the pairwise \textbf{Spearman correlation coefficients} among the main visual metrics aggregated over all cities. Correlations were computed on standardized 1–5 scores after excluding missing or invalid entries. The resulting coefficients highlight the structural coupling and independence between key dimensions of urban visual quality.

\begin{table}[htbp]
    \centering
    \small
    \caption{Spearman correlation coefficients among structural and contextual visual metrics (all cities combined, $N{=}3000$).}
    \label{tab:correlation_matrix}
    \begin{tabular}{lrrrrrr}
        \toprule
        & \textbf{Pavement} & \textbf{Façade} & \textbf{Greenery} & \textbf{Graffiti} & \textbf{Canyon} & \textbf{Material} \\
        \midrule
        \textbf{Pavement (PCI)} & 1.00 & 0.22 & $-0.08$ & $-0.03$ & $-0.05$ & 0.02 \\
        \textbf{Façade (FDS)} & 0.22 & 1.00 & 0.35 & $-0.05$ & 0.13 & $-0.00$ \\
        \textbf{Greenery} & $-0.08$ & 0.35 & 1.00 & 0.04 & 0.30 & $-0.04$ \\
        \textbf{Graffiti} & $-0.03$ & $-0.05$ & 0.04 & 1.00 & 0.00 & 0.04 \\
        \textbf{Canyon} & $-0.05$ & 0.13 & 0.30 & 0.00 & 1.00 & $-0.06$ \\
        \textbf{Material} & 0.02 & $-0.00$ & $-0.04$ & 0.04 & $-0.06$ & 1.00 \\
        \bottomrule
    \end{tabular}
    \vspace{1ex}
\end{table}

\noindent
The correlation structure reveals several interesting patterns:

\begin{itemize}
    \item \textit{Pavement and Façade Quality ($\rho = 0.22$):} A weak positive correlation indicates that better road conditions are often found alongside better-maintained façades, suggesting that maintenance efforts may cluster spatially.
    \item \textit{Façade Quality and Greenery ($\rho = 0.35$):} The strongest, but still moderate, observed correlation highlights that greener streetscapes are generally associated with higher façade quality, reinforcing the link between environmental and visual well-being.
    \item \textit{Façade Quality and Graffiti ($\rho = -0.05)$:} The expected negative relationship suggests that graffiti presence slightly co-varies with facade degradation, although the effect remains very weak, almost negligible.
    \item \textit{Urban Canyon and Greenery ($\rho = 0.30$):} Denser built-up areas tend to include more vegetation in visible frames, possibly due to the presence of trees along narrow streets, a configuration typical of Mediterranean city cores.
    \item \textit{Material and Pavement Quality ($\rho = 0.02$):} The absence of a clear relationship confirms that the type of pavement material is not directly predictive of its visual state at the scale of analysis considered.
\end{itemize}

\noindent
Overall, the correlation magnitudes remain quite modest ($|\rho| < 0.4$), confirming that the various visual metrics capture \textit{distinct and complementary dimensions} of urban quality. While Pavement and Façade scores show some shared structure, contextual variables such as Greenery, Graffiti, and Urban Canyon contribute independent information to the city’s morphological and perceptual profile.

\begin{figure}[htbp]
    \centering
    \includegraphics[width=0.75\textwidth]{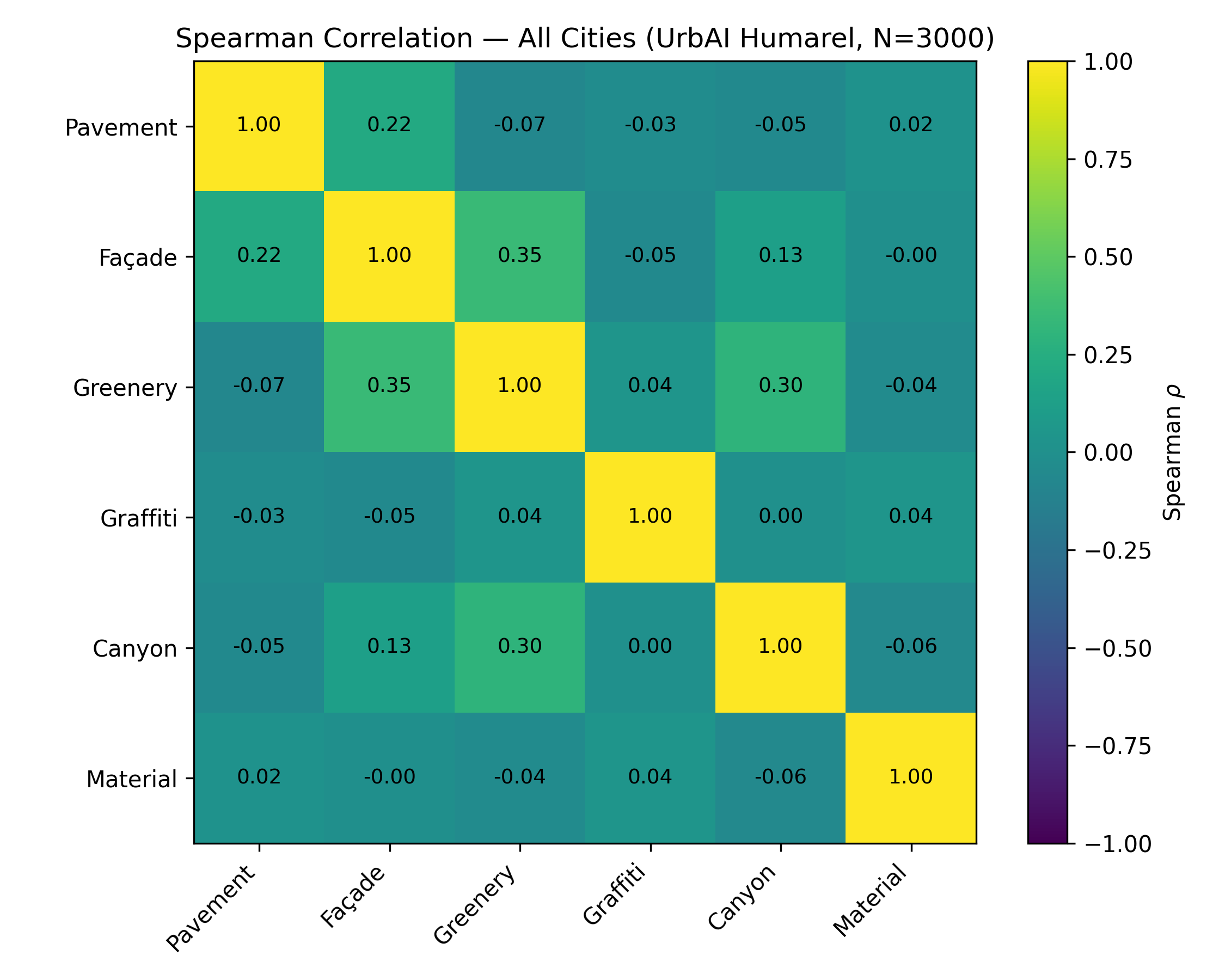}
    \caption{Spearman correlation heatmap of visual metrics across all cities. Colors range from blue (negative correlation) to red (positive correlation). The strongest relationship appears between Façade and Greenery, while other pairs show weaker, complementary dependencies.}
    \label{fig:correlation_heatmap}
\end{figure}

Figure~\ref{fig:correlation_heatmap} provides a visual overview of these relationships, illustrating the moderate positive association between Façade and Greenery scores and the weak coupling of the remaining indicators. The relative independence of these metrics supports their combined use in a multivariate modeling framework for large-scale urban analysis to process expanded datasets across multiple European cities.

\begin{table}[htbp]
    \centering
    \small
    \caption{Spearman correlation coefficients among structural and contextual visual metrics (all cities combined, $N{=}3000$).}
    \label{tab:correlation_matrix}
    \begin{tabular}{lrrrrrr}
        \toprule
        & \textbf{Pavement} & \textbf{Façade} & \textbf{Greenery} & \textbf{Graffiti} & \textbf{Canyon} & \textbf{Material} \\
        \midrule
        \textbf{Pavement (PCI)} & 1.00 & 0.22 & $-0.08$ & $-0.03$ & $-0.05$ & 0.02 \\
        \textbf{Façade (FDS)} & 0.22 & 1.00 & 0.35 & $-0.05$ & 0.13 & $-0.00$ \\
        \textbf{Greenery} & $-0.08$ & 0.35 & 1.00 & 0.04 & 0.30 & $-0.04$ \\
        \textbf{Graffiti} & $-0.03$ & $-0.05$ & 0.04 & 1.00 & 0.00 & 0.04 \\
        \textbf{Canyon} & $-0.05$ & 0.13 & 0.30 & 0.00 & 1.00 & $-0.06$ \\
        \textbf{Material} & 0.02 & $-0.00$ & $-0.04$ & 0.04 & $-0.06$ & 1.00 \\
        \bottomrule
    \end{tabular}
    \vspace{1ex}
\end{table}

\subsection{City-Level Correlation Patterns}

To verify whether the relationships observed in the global correlation matrix are consistent across different urban morphologies, we computed the pairwise Spearman coefficients for each of the six metropolitan areas (Table~\ref{tab:city_corr_summary}). The analysis focuses on four key relationships: Pavement–Façade, Façade–Greenery, Façade–Graffiti, and Greenery–Canyon.

\begin{table}[htbp]
    \centering
    \small
    \caption{Spearman correlation coefficients for key metric pairs computed separately for each city.}
    \label{tab:city_corr_summary}
    \begin{tabular}{lrrrr}
        \toprule
        \textbf{City} & \textbf{$\rho$(Pavement, Façade)} & \textbf{$\rho$(Façade, Greenery)} & \textbf{$\rho$(Façade, Graffiti)} & \textbf{$\rho$(Greenery, Canyon)} \\
        \midrule
        Bologna & 0.10 & 0.27 & $-0.05$ & \textbf{0.41} \\
        Florence & $-0.04$ & \textbf{0.45} & $-0.07$ & 0.19 \\
        Milan & \textbf{0.27} & 0.21 & $-0.09$ & 0.27 \\
        Naples & 0.26 & \textbf{0.44} & \textbf{$-0.13$} & 0.32 \\
        Palermo & 0.20 & \textbf{0.42} & $-0.01$ & \textbf{0.38} \\
        Rome & 0.24 & 0.24 & $-0.05$ & 0.23 \\
        \bottomrule
    \end{tabular}
    \vspace{1ex}
\end{table}

\noindent
The results confirm that the overall structure of correlations is robust across cities, with all coefficients remaining within the mild-to-moderate range ($|\rho| < 0.5$). The \textit{Façade–Greenery} relation emerges as the most stable and substantial, reaching $\rho \approx 0.45$ in Florence and Naples, while \textit{Pavement–Façade} correlations are weaker ($\rho \approx 0.10$–0.27) but consistently positive. \textit{Façade–Graffiti} correlations remain negative in all cases, indicating that higher graffiti presence is associated with lower façade quality, with Naples showing the strongest effect ($\rho = -0.13$).

\noindent
Taken together, these city-level results confirm the general trends observed in the aggregated analysis while revealing distinct local signatures. The persistence of consistent correlation signs across diverse morphologies underscores the stability of the UrbIA visual metrics, while the magnitude variations highlight how local planning histories modulate the coupling between infrastructural and contextual dimensions of urban quality.

\section{Discussion and Conclusions}

The results presented in this paper provide a coherent, data-driven portrait of the visual and infrastructural complexity of six major Italian cities. 
Across all metrics, the analysis reveals a high degree of spatial heterogeneity, confirming that visual quality is not evenly distributed but rather fragmented into a patchwork of well-maintained and degraded areas. 
The weak or statistically insignificant \textit{Urban Gradients} observed (typically $R^2 < 0.03$) indicate that distance from the historical core alone no longer explains variations in infrastructure quality, supporting the notion that contemporary Italian cities are intrinsically polycentric and spatially stratified. 
The \textit{Cross-Metric Correlation Analysis} complements this picture by showing modest but consistent associations among visual indicators, particularly the positive link between façade quality and greenery ($\rho \approx 0.35$), suggesting that structural and environmental qualities co-vary in weak yet interpretable ways.

From a morphological perspective, these findings imply that Italian metropolitan areas exhibit a high level of internal differentiation: the coexistence of well-preserved heritage districts, recently renewed zones, and neglected interstitial spaces. 
Such complexity challenges traditional models of concentric urban decay and supports a view of the city as a multi-scalar, heterogeneous system shaped by successive waves of construction, conservation, and infrastructural maintenance. 
The strong intra-urban contrast revealed by the \textit{Spatial Variance} measures may thus be interpreted as an empirical signature of this historical layering and functional diversification.

Methodologically, this study demonstrates the viability of large-scale, image-based urban sensing using multimodal Vision Intelligence. 
Despite the presence of approximately 15\% visually ambiguous or low-quality images—an inherent limitation of Street View sampling—the indicators defined here prove statistically robust, producing stable distributions and correlations across all cities. 
Future refinements will focus on improving the \textit{Humarel} agents through better tuning of visual recognition parameters, adaptive frame selection, and automated filtering of anomalous or non-representative images. 
Enhanced calibration and multi-view consistency checks will allow the extraction of more precise and semantically consistent visual metrics.

TThe next phase of this work, currently under development, involves scaling the 
analysis to national and European levels using cloud-based GPU infrastructure for 
large-scale image processing and vision model inference.
This expansion will enable the processing of millions of Humarel observations and the integration of additional contextual layers, such as socioeconomic indicators, mobility patterns, energy performance, and land-use statistics. 
By correlating visual metrics with these external data sources, we aim to explore how physical appearance, environmental quality, and social conditions intertwine within complex urban systems. 
Ultimately, the project will transition from descriptive analysis to predictive modeling, employing statistical learning and simulation to anticipate spatial dynamics of degradation, maintenance, and renewal. 

Beyond its empirical findings, this study illustrates how \textit{Urban Vision Intelligence} can serve as a foundational tool for quantitative morphology and policy-oriented diagnostics, bridging visual perception, infrastructure assessment, and computational urban science. 

\vspace{6pt}

\section*{Author Contributions}
Conceptualization, M.D.E.; methodology, M.D.E. and M.F.; 
software, M.D.E., M.F., and M.M.; 
writing—original draft preparation, M.D.E.; 
writing—review and editing, M.D.E., A.B., M.P., M.M., M.F., and C.D.C. 
All authors have read and agreed to the published version of the manuscript.

\section*{Funding}
This research received no external funding.

\section*{Institutional Review Board Statement}
Not applicable.

\section*{Informed Consent Statement}
Not applicable.

\section*{Data Availability Statement}
The datasets generated and analyzed during the current study are available from the corresponding author upon request.


\section*{Conflicts of Interest}
The authors declare no conflicts of interest.


\section*{Abbreviations}
The following abbreviations are used in this manuscript:

\begin{center}
\begin{tabular}{@{}ll@{}}
AI  & Artificial Intelligence \\
FDS & Façade Degradation Score \\
GMP & Google Maps Platform \\
GSV & Google Street View \\
LLM & Large Language Model \\
PCI & Pavement Condition Index \\
UVI & Urban Visual Intelligence \\
ViT & Vision Transformer
\end{tabular}
\end{center}





\end{document}